# Effects of non-magnetic defects in hole doped cuprates: exploration of the roles of the underlying electronic correlations

S. H. Naqib* and R. S. Islam

*Department of Physics, University of Rajshahi, Raj-6205, Bangladesh*

## Abstract

The effects of non-magnetic iso-valent defects inside the $CuO_2$ plane(s) on superconducting transition temperature, $T_c$, dc charge transport, and the bulk magnetic susceptibility, $\chi(T)$, were investigated for the $YBa_2(Cu_{1-y}Zn_y)_3O_{7-\delta}$ (Zn-YBCO) and $La_{2-x}Sr_xCu_{1-y}Zn_yO_4$ (Zn-LSCO) superconductors over a wide range of hole concentrations, $p$, and Zn contents ($y$) in the $CuO_2$ plane(s). From the analysis of the $\chi(T, y)$ data, the pseudogap energy scale, $\varepsilon_g$, was found to be almost independent of the defect content at a given value of $p$. The Zn induced rate of suppression of $T_c$, $dT_c(p)/dy$, was found to be strongly $p$-dependent and showed a systematic variation with hole concentration, except in the vicinity of $p \sim 0.125$, *i.e.*, near the so-called $1/8^{th}$ anomaly where the charge and spin stripe orderings are at their strongest in various families of hole doped cuprates. Near $p \sim 0.125$, the static striped charge ordering is largely believed to dominate the $T$-$p$ electronic phase diagram. $dT_c(p)/dy$ decreased strongly around this composition, *i.e.*, Zn abruptly became less effective in degrading $T_c$ when $p \sim 0.125$. This observation, together with the facts that (i) Zn suppresses $T_c$ most effectively and (ii) the characteristic pseudogap energy scale remains insensitive to the level of Zn substitution, provide us with important clues regarding the nature and interplay of the underlying electronic correlations present in high-$T_c$ cuprates. We have discussed the possible connections among the stripe ordering, superconducting correlations and the pseudogap phenomenon in details in view of the above findings.

# 1. Introduction

It is known since the early days of cuprate superconductors that high-$T_c$ superconductivity emerges while charge carriers ($p$) are added in the $CuO_2$ planes of the parent antiferromagnetic insulating (AFMI) compounds. The $T$-$p$ phase diagram exhibits a number of interesting features arising from the strongly correlated electronic nature of these materials as the number of doped carriers, $p$, increases, *e.g.*, AFMI, spin glass (SG), non Fermi-liquid (FL) superconducting (SC), strongly spin/charge ordered (striped), and more conventional (FL-like) SC states [1 – 7]. The possible interplay among these ground states and their relevance to superconductivity are a matter of intense debate. The complete understanding of the physics of cuprate superconductivity and the various normal state correlations remains one of the fundamental issues in condensed matter physics, still after more than quarter of a century of their discovery. It is widely believed that an understanding of these correlations in the normal state is the essential prerequisite for developing a successful microscopic theory of the superconducting state of high-$T_c$ cuprates.

Besides superconductivity itself, PG remains the most extensively studied phenomenon in hole doped cuprates. The PG is detected by different experimental probes over a certain range of hole content, extending from the underdoped (UD) to the slightly overdoped (OD) regions [1 - 3]. In the PG region, a number of non-Fermi-liquid like features are observed both in normal and SC states where contrary to one of the central tenants of the Landau quasi-particle (QP) picture, low-energy electronic excitations are gapped along certain directions of the Brillouin zone while *Fermi-arcs* survive in other directions [3]. Existing theoretical proposals to explain the origin and the evolution of the PG with hole content and temperature could be classified broadly into two groups. One is based on the preformed pairing scenario, where PG arises from strong fluctuations of SC origin in the strong coupling limit for systems with low dimensionality (high structural and physical anisotropy) and low superfluid density [4]. Within the second framework, PG is attributed to some correlations of non-SC origin. Here PG is thought to coexist and, in fact, competes with superconductivity [1]. Considerable debate has ensued as to the nature of the PG and no consensus has been reached [1, 2]. On the other hand, the static spin/charge stripe correlations are only observed in the UD cuprates in the vicinity of $p \sim$ 0.125 (the so-called $1/8^{th}$ anomaly) [6, 7], although dynamical (fluctuating) charge/spin stripe correlations are believed to exist over a much wider doping range, especially in the La214 compounds [6, 7]. Incommensurate low-energy spin fluctuations, generally interpreted as the precursor to stripe correlations, are also observed in Y123 and Bi2212 [6 – 8]. Therefore, there is reason to believe that tendency for stripe ordering is a generic feature of all hole doped cuprate superconductors. Stripe phase probably forms in doped Mott insulators as a compromise between the AFM ordering among the Cu spins and strong Coulomb interaction between the electrons (both favoring localization) with the kinetic energy of the mobile charge carriers (leading to delocalization). Broadly speaking, stripe phase can be viewed as spontaneously separated ordered states of charge-rich (high kinetic energy) and charge-poor (antiferromagnetically correlated) regions throughout the compound. Stripes can be both one-dimensional (1D) and two-dimensional (2D) in nature. The significance of spin/charge ordered states in the physics of hole doped cuprates have been demonstrated further by recent comparative study of the thermoelectric transports in Y123 and Eu-LSCO systems [9]. This work [9] boldly suggests that the Fermi-surface reconstruction, as seen by quantum oscillation experiments [10] at low temperatures, is possibly due to the breaking of the lattice translational symmetry caused by the stripe ordering. The appearance of field induced charge ordering in Y123 [11] adds strong support to the assumption that stripe order is an essential ingredient for the understanding of these strongly correlated electronic systems. The thermoelectric transport properties also showed that the basic features of the stripe related phase diagrams for Y123 and Eu-LSCO are essentially the same and the charged stripe phase exists over quite an extended doping range from $p = 0.08$ to 0.18 [9] in both the families. It is somewhat agreed that static stripe order degrades superconductivity [5 – 7, 9], but the possible influence of fluctuating stripe order on superconductivity is a matter of significant interest [6, 7]. Some of the theoretical models link the origins of both superconductivity and PG to the stripe correlations [6, 7, 12 – 14]. A proper understanding of the nature of the interplay among the SC, PG, and stripe correlations are therefore essential to construct a coherent theoretical picture explaining the normal and superconducting states of hole doped cuprates.

Substituting a single impurity atom for an in-plane Cu strongly perturbs the surrounding electronic environment and can be used as a probe for understanding the physics of high-temperature superconductors. Therefore, the study of the effects of controlled in-plane disorder on the various electronic correlations in cuprates has the potential to yield valuable information regarding their nature and possible interrelations. $Zn^{2+}$ is a non-magnetic ion owing to its $3d^{10}$ (spin zero, $s = 0$) electronic configuration, but it still is a very strong destroyer of superconductivity. Zn causes a rapid reduction of $T_c$, of the order of ~ 10 K/%Zn in the $CuO_2$ plane [15, 16]. Another interesting result of non-magnetic Zn substitution is the defect-induced enhanced magnetism. Various experiments [17 – 19] gave *indications* that iso-valent Zn gives rise to a localized magnetic moment-like feature on the four nearest neighbor Cu sites in the $CuO_2$ planes. This apparent Zn induced magnetic behavior is evident from the appearance of the Curie-like term in the bulk magnetic susceptibility, $\chi(T)$, in the normal state [17, 18]. Whether this enhanced susceptibility is truly due to localized moments in the immediate neighborhood of Zn atoms [20] or due to a redistribution of the QP spectral weight in the presence of a PG [21] is a matter yet to be settled. On the other hand, in-plane defects have the potential to stabilize dynamical stripe order (via *stripe pinning*) [6, 7]. Therefore, Zn substituted hole doped cuprates provide us with a unique opportunity to study the intricate interplay among SC, PG, and stripe correlations.

The effect of Zn on the PG has been studied before, mainly by charge transport [22 – 24] and electronic heat capacity measurements [25]. In this article, we present a model for the bulk magnetic susceptibility mainly in terms of a depleted electronic density of states at low-energy due to the PG and a Zn induced enhancement resembling to a Curie contribution. This simple model describes the experimental data quite well and yields an average (bulk) value of the magnitude of the PG energy scale as a function of hole and disorder contents. From the observed $p$-dependences of the enhanced magnetic susceptibility, it appears that Zn induced magnetic behavior and the magnitudes of the PG are closely related. The PG energy scale does not vary with Zn, even in samples where $T_c$ has been completely suppressed by disorder. In this article, we have reported some comprehensive results of charge transport and bulk magnetic susceptibility measurements of single layer $La_{2-x}Sr_xCu_{1-y}Zn_yO_4$ and double-layer $YBa_2(Cu_{1-y}Zn_y)_3O_{7-\delta}$ systems. From the analysis of the data and the observed $p$ dependences of $T_c$, $\varepsilon_g$, and $dT_c/dy$ (the rate of suppression of $T_c$ with Zn) point towards a scenario where both PG and stripe correlations are not directly related to fluctuation superconductivity. The strong stripe ordering in the vicinity of the $1/8^{th}$ anomaly affects the rate of suppression of $T_c$ in a non-trivial fashion but appears to have a minimal effect on $\varepsilon_g(p)$.

## 2. Experimental samples and their characterizations

Phase pure polycrystalline $La_{2-x}Sr_xCu_{1-y}Zn_yO_4$ compounds were synthesized by solid state reaction method using the following powdered chemicals from *ALDRICH* (purity is given in brackets): $La_2O_3$ (99.999%), $SrCO_3$ (99.995%), CuO (99.9999%) and ZnO (99.999%). All these compounds were dried in alumina crucibles in a furnace at 500°C ($SrCO_3$ in a separate furnace at 400°C) for 10 hours so as to eliminate any water absorbed (which could lead to stoichiometric errors) in the chemicals. The dried powders were weighed in proper stoichiometric proportions (0.02 mole of each to give ~ 8 gm of the desired compound) and ground together in an agate mortar under cyclohexane (99.94%) to help the homogeneous mixing of the powders. These loose powdered mixtures were first calcined at 920°C in air for 12 hours (heating/cooling rate = 300°C/hour). The purpose of this step was to decompose $SrCO_3$ into SrO and $CO_2$ and thus eliminate the unwanted $CO_2$. To check whether the $SrCO_3$ had decomposed properly, the powders were weighed before and after the calcination process. A weight loss of approximately 1% was to be expected. The powders were reground and the above calcination process was repeated, this time with dwell period of 24 hours, to check whether there were any further weight losses. Quite small weight losses were found (between 0.19% - 0.35%) at this stage. The powders were then reground, pressed into pellets (with diameters ~ 15 mm) under ~ 6 ton/cm$^2$ pressure and sintered at 985°C for 12 hours in a tube furnace under air flow (high purity 'control air'), with an intermediate step at 920°C (for 10 hours). This sintering procedure was repeated once more. XRD was done after each sintering. The densities obtained (4.8 - 5.1 gm/cm$^3$) were lower than desired. So the samples were reground and pressed into pellets for the final sintering at 1010°C for 12 hours, with an intermediate step at 985°C (for 12 hours). The final densities were 6.09 - 6.29 gm/cm$^3$

These values were higher than that found by Fisher *et al*. (5.6 gm/cm$^3$) [26] for the same sintering temperature. From published lattice parameters [27], the theoretical X-ray density is ~ 7.0 gm/cm$^3$. As far as dc charge transport is concerned, density is an important parameter, since high density sintered compounds show better grain connectivity and therefore, a lower resistivity.

Sintered samples of YBa$_2$(Cu$_{1-y}$Zn$_y$)$_3$O$_{7-\delta}$ were prepared following the method described in refs. [28, 29]. These samples were also used as target materials for the thin films grown by the pulsed laser ablation process. Polycrystalline samples were synthesized using the following powdered chemicals supplied by *SIGMA-ALDRICH* (purity is given in brackets): Y$_2$O$_3$ (99.999%), BaCO$_3$ (99.999%), CuO (99.9999%), and ZnO (99.999%). All these chemicals were dried in alumina (Al$_2$O$_3$) crucibles in a furnace at 500°C in air for 12 hours. The dried powders were weighed in the proper stoichiometric ratios and ground together in an agate mortar with cyclohexane (99.94%) to help a homogeneous mixing of the powders. These well-ground powdered mixtures were then put into a box furnace at 150°C for further drying. The loose powdered mixtures were first calcined at 900°C in air for 12 hours (heating/cooling rate = 300°C/hour). This procedure was repeated thrice (with intermediate grindings at each step) but at temperatures 910°C, 920°C, and 935°C respectively. Pellets were formed using a pressure of 7 tons/cm$^2$ and the 935°C (in air) run was repeated once more. After each run we have noted the mass change. Significant weight losses were observed after the first sintering of the pellets at 935°C, all our samples lost ~ 12% of their initial masses. This we believe was due to nearly the complete loss of unwanted CO$_2$. Sintering runs above 935°C were performed in high-purity (*5-Grade*) O$_2$ at a pressure of 1 bar. The sintering temperatures were increased stepwise (with intermediate grindings) by 10°C from 960°C to 980°C (kept 12 hours at each temperature). Samples were furnace-cooled to room-temperature after each sintering in O$_2$. We have used a gas purification system during all the sintering runs which removed possible presence of moisture and traces of CO$_2$ in the gases used. The evolution of various phases in the sample with different calcination conditions were tracked by X-ray diffraction (XRD) on both powders and pellets. After the 980°C sintering in O$_2$, all of the compounds appeared almost completely phase-pure, few of them showed minute traces of impurity phases at levels < 1.5% of the XRD count. Representative XRD data for $La_{2-x}Sr_xCu_{1-y}Zn_yO_4$ and $YBa_2(Cu_{1-y}Zn_y)_3O_{7-\delta}$ compounds are shown in Figs. 1 – 3.

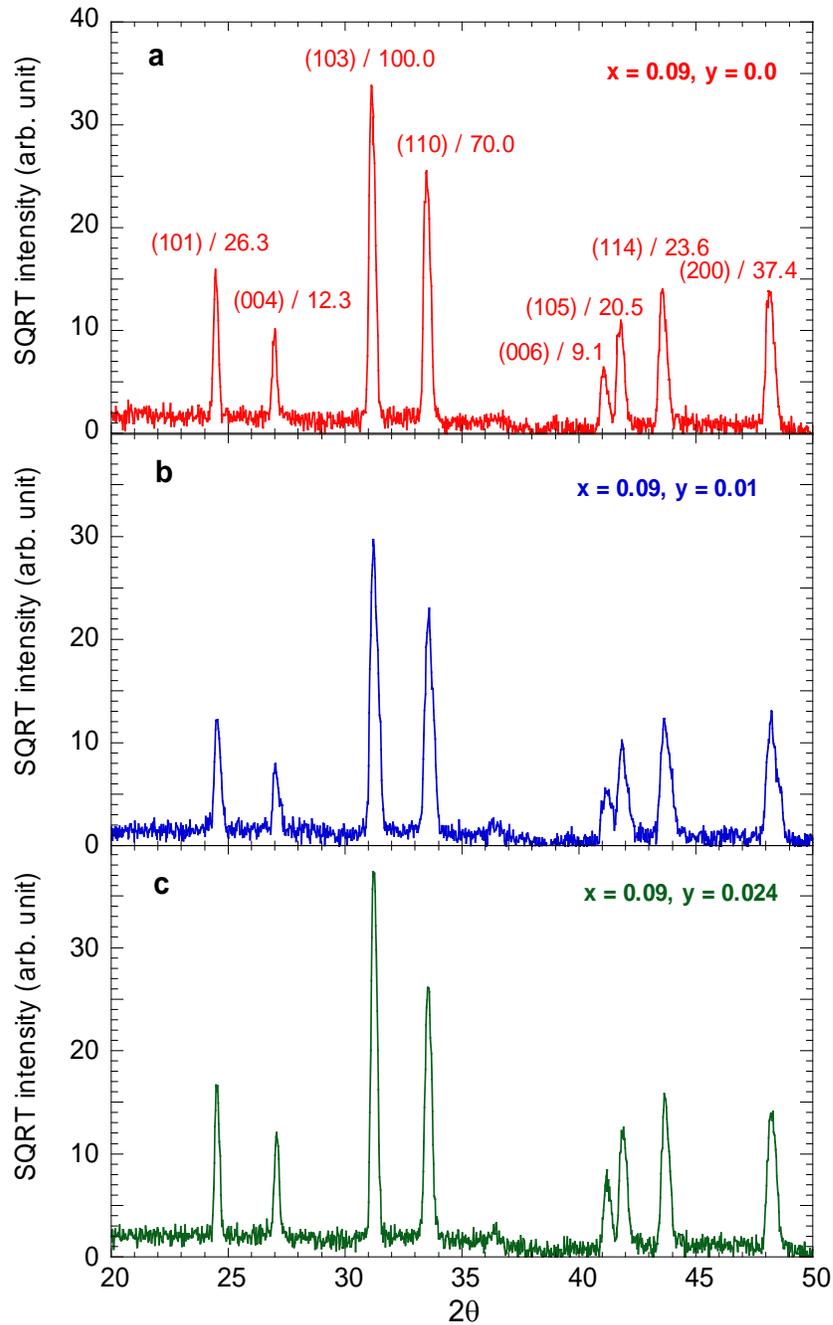

Fig. 1 XRD plots for $La_{2-x}Sr_xCu_{1-y}Zn_yO_4$ (sintered pellets), $x$ and $y$-values are given in the figures. Miller indices and relative intensities are shown for the Zn-free compound.

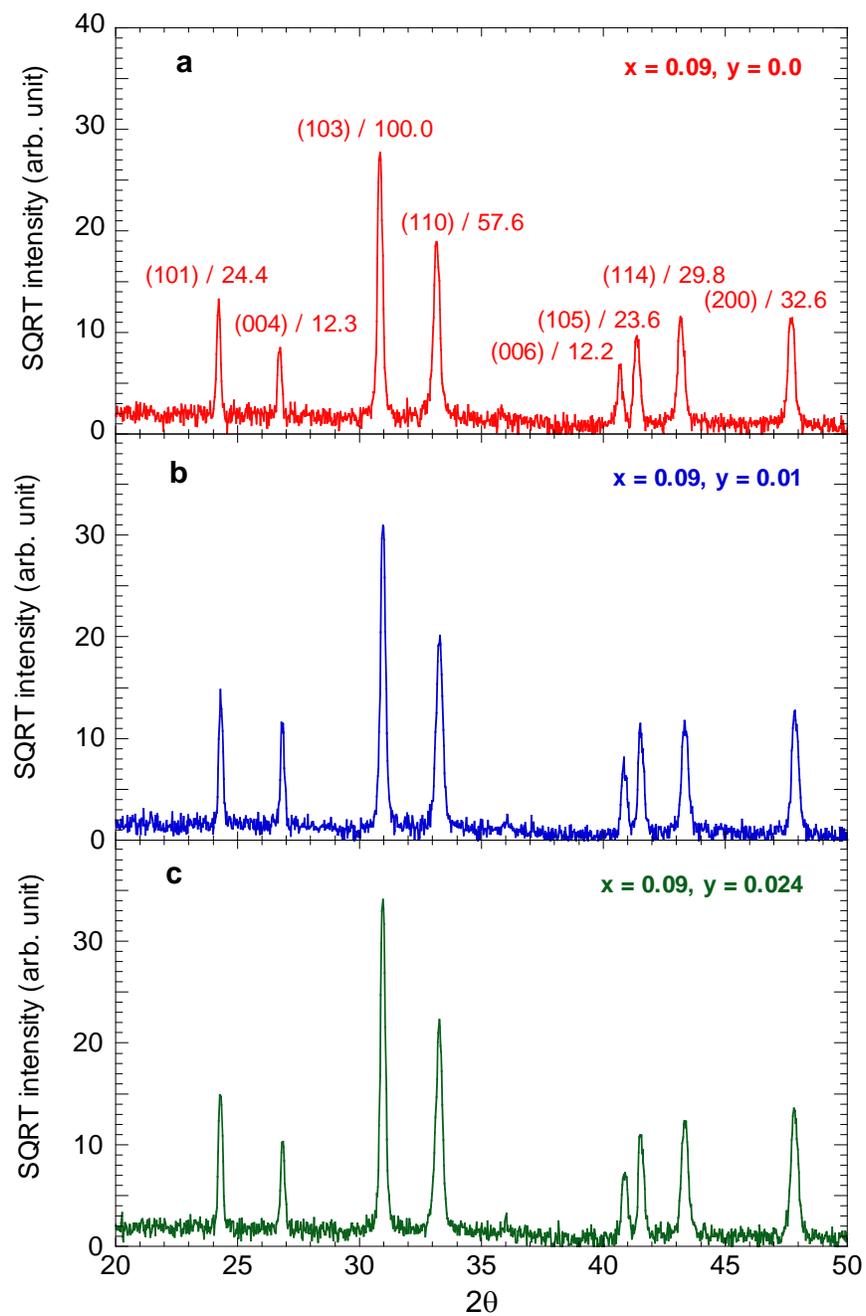

Fig. 2 XRD plots for $La_{2-x}Sr_xCu_{1-y}Zn_yO_4$ (powder), $x$ and $y$-values are given in the figures. Miller indices and relative intensities are shown for the Zn-free compound.

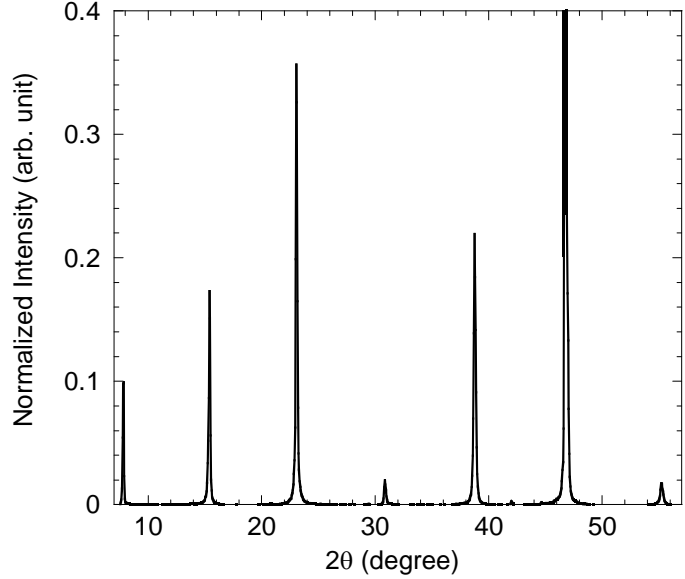

Fig. 3   XRD plot for an optimally doped YBa$_2$(Cu$_{1-y}$Zn$_y$)$_3$O$_{7-\delta}$ thin film. Only the (00$l$) peaks are visible due to epitaxial growth.

In this article we report results obtained for La$_{2-x}$Sr$_x$Cu$_{1-y}$Zn$_y$O$_4$ with the following compositions: Sr content $x$ (= $p$, for La$_{2-x}$Sr$_x$Cu$_{1-y}$Zn$_y$O$_4$) = 0.08 (Zn-free only), 0.09, 0.10, 0.11, 0.12, 0.14, 0.15, 0.17 (Zn-free only), 0.18 (Zn-free only), 0.19, 0.22, 0.27 (Zn-free only) and $y$ = 0.0, 0.005, 0.01, 0.015, 0.02, 0.024. The hole content of the Zn-substituted Y123 compounds were varied by oxygen annealing at different temperatures and gas pressures. The Y123 compounds were grown with compositions of $y$ = 0.0, 0.02, 0.03, and 0.04. All these samples were characterized by XRD and room-temperature thermopower ($S$[290 K]). $S$[290 K] gave an accurate measure of the hole content in the CuO$_2$ plane(s) irrespective of disorder content and the crystalline state [30]. $T_c$ for sintered La$_{2-x}$Sr$_x$Cu$_{1-y}$Zn$_y$O$_4$ and YBa$_2$(Cu$_{1-y}$Zn$_y$)$_3$O$_{7-\delta}$ compounds were determined from low-field ($H$ = 1 Oe, $f$ = 333.33 Hz) AC susceptibility (ACS) and resistivity, $\rho(T)$, measurements. High-quality $c$-axis oriented crystalline thin films of YBa$_2$(Cu$_{1-y}$Zn$_y$)$_3$O$_{7-\delta}$ were fabricated using the pulsed LASER deposition (PLD) method. Details of thin film fabrication can be found elsewhere [29, 31]. For these epitaxial thin films of YBa$_2$(Cu$_{1-y}$Zn$_y$)$_3$O$_{7-\delta}$, $T_c$ was determined from the in-plane resistivity, $\rho_{ab}(T)$, measurements only. Thickness and surface morphology of the thin films were studied using atomic force microscopy (AFM). Thickness of all the films lie within (2800 ± 400) Å. Films grown under optimal conditions exhibited high surface quality and large single crystalline grains. The maximum $T_c$ of the thin films were found to be consistently lowered by ~ 2 K as compared to the $T_c$ for sintered samples at the same hole concentration. A possible reason for this might be the non-uniform epitaxial strain due to the lattice mismatch between the target material and the SrTiO$_3$ substrates. Alternatively it could be caused by unintentional in-plane atomic disorder present in the films.

## 3. Experimental results and analysis

ACS was measured, using a commercial *Lake Shore Cryotronics* Model 7000 AC susceptometer, both on bar shaped sintered compound (with magnetic field applied along the longest dimension) and on powder. The AC susceptometer was calibrated using a Pb sphere. When completely diamagnetic, such a sphere yields a signal of 23.9 $\mu V$/mm$^3$ (with $H$ = 1 Oe and $f$ = 333.33 Hz). Most of the samples used in this study exhibited sharp SC transitions (with a transition width $\Delta T_c$ < 0.5 K). Transition widths increase somewhat for the heavily Zn substituted compounds. ACS results for some of the sintered La$_{2-x}$Sr$_x$Cu$_{1-y}$Zn$_y$O$_4$ compounds are shown in Figs. 4. $T_c$ was determined keeping the resolution of the susceptometer and the effects of Gaussian fluctuation diamagnetism in mind as follows: a straight line was drawn at the steepest part near the onset of the diamagnetic ACS curve (where the magnitude of the ACS signal does not exceed a few microvolts. This follows from the estimated orders of magnitude of the contribution to the ACS signal originating from diamagnetic fluctuations just above $T_c$ for La214 and Y123 [32]), and another one was drawn as the $T$-independent base line associated with negligibly small normal state signal. The intercept of the two lines gave $T_c$. A typical example is shown in Fig. 5. This procedure yields almost identical (~ 1 K higher for the powdered compounds) values of $T_c$ both for sintered and powdered samples.

Patterned thin films with evaporated gold contact pads and high density (89 to 93% of the theoretical density) sintered bars were used for resistivity measurements. Resistivity was measured using the four-terminal method with an ac current of 1 mA (77 Hz), using 40 μm dia. copper wire and silver paint to make the low resistance contacts. From resistive measurements, $T_c$ was taken at zero resistance transition point (within the noise level of ± 10$^{-6}$ Ω). An example, illustrating this method of determining $T_c$ is shown in Fig. 6. We have also shown the ACS data for the same sample in Fig. 6. Representative plots of resistivity data for films and sintered YBa$_2$(Cu$_{1-y}$Zn$_y$)$_3$O$_{7-\delta}$ are shown in Figs. 7.

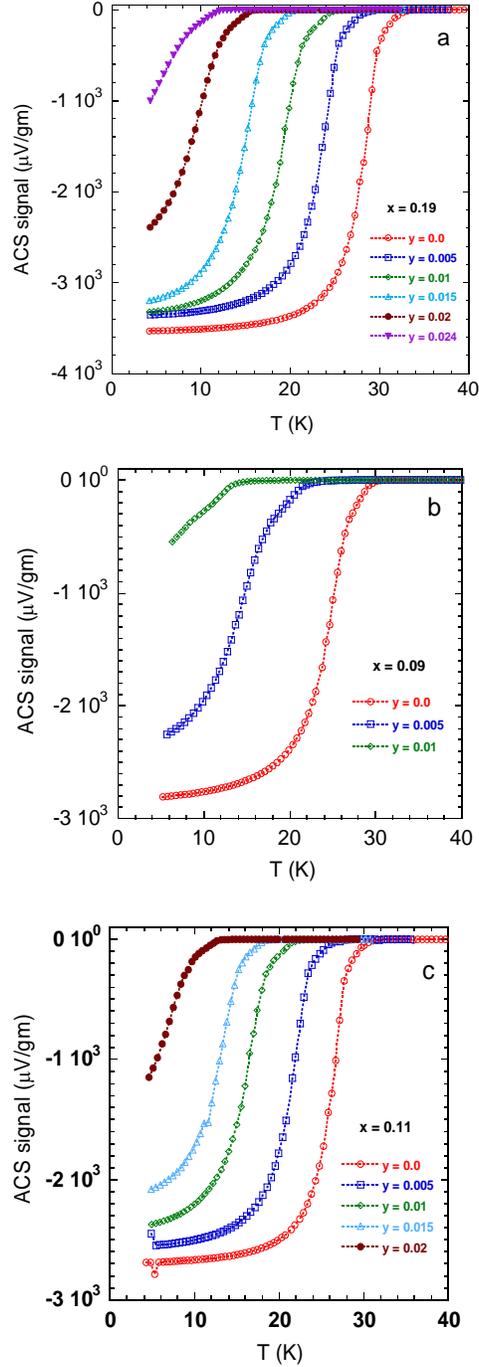

Fig. 4  ACS results for some representative sintered La$_{2-x}$Sr$_x$Cu$_{1-y}$Zn$_y$O$_4$ samples (a) $x = 0.19$ (OD), (b) $x = 0.09$ (UD), and (c) $x = 0.11$ (UD, near the 1/8$^{th}$ anomaly) compounds.

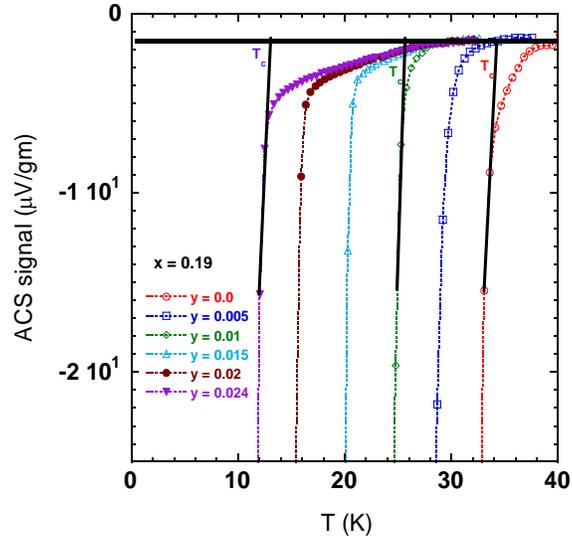

Fig. 5 Determination of $T_c$ from the ACS data (see text for details).

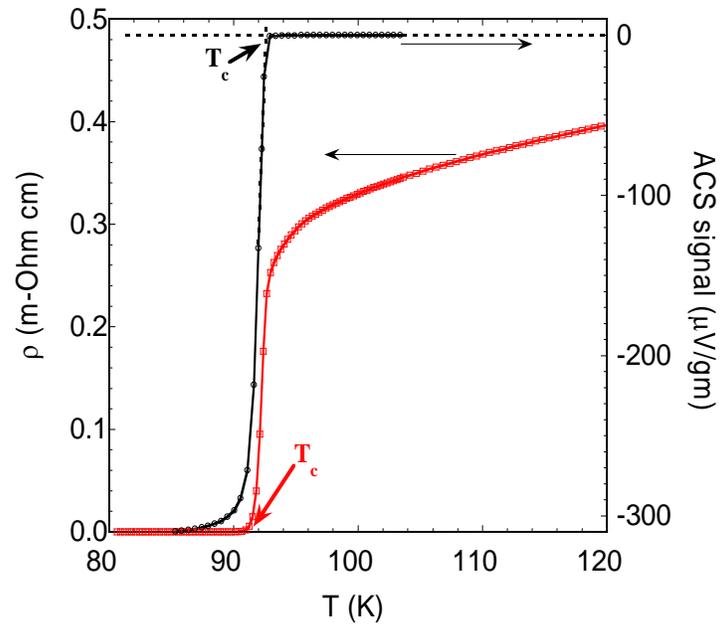

Fig. 6 Determination of $T_c$ from the low-field ACS (black circles) and the resistivity data (red squares) for an optimally doped pure Y123 ($p = 0.160 \pm 0.004$) sintered compound.

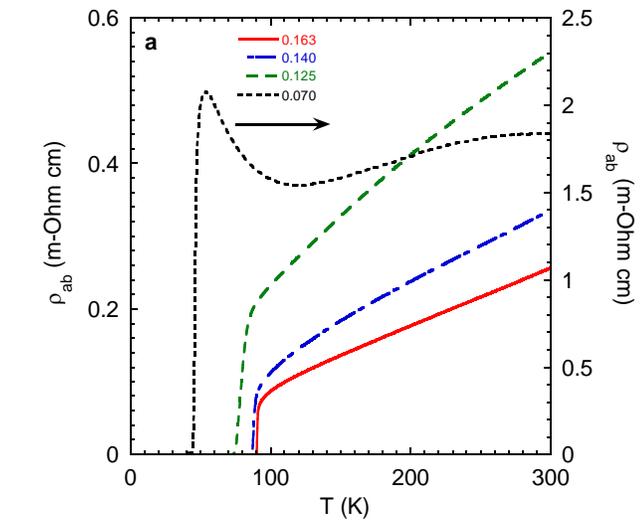

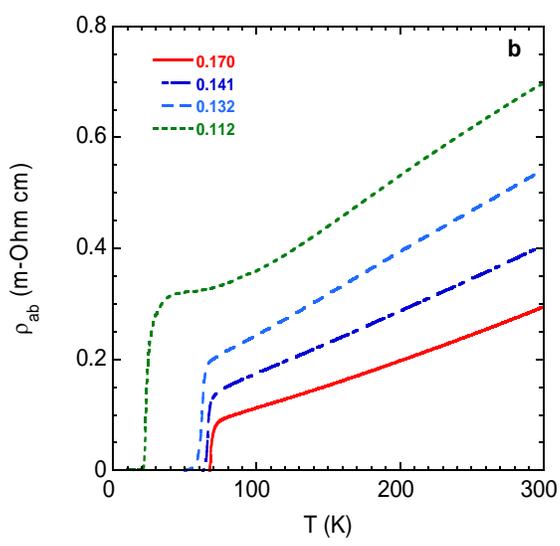

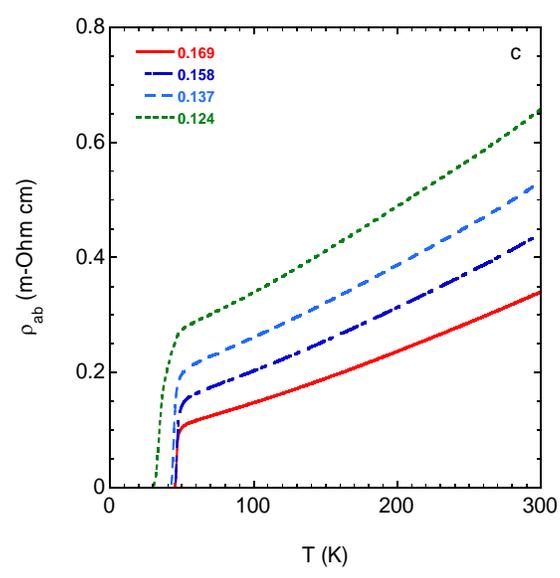

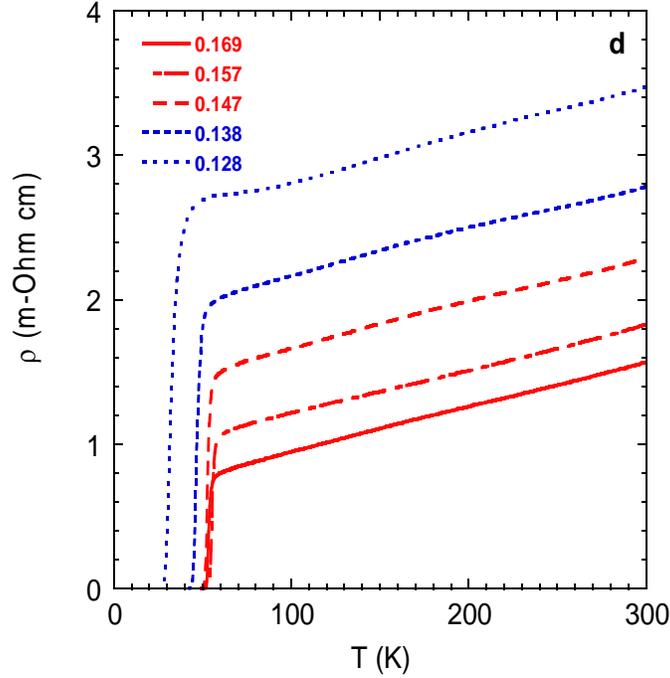

Fig. 7 In-plane resistivity, $\rho_{ab}(T)$, for some representative c-axis oriented crystalline thin films of $YBa_2(Cu_{1-y}Zn_y)_3O_{7-\delta}$ with (a) $y = 0.00$, (b) $y = 0.02$, (c) $y = 0.04$, and (d) Resistivity data for $y = 0.03$ sintered Y123 compounds. The hole contents (within $\pm 0.004$) are given in the plots.

The bulk static magnetic susceptibility, $\chi(T)$, measurements on sintered $La_{2-x}Sr_xCu_{1-y}Zn_yO_4$ and $YBa_2(Cu_{1-y}Zn_y)_3O_{7-\delta}$ compounds were performed using a *Quantum Design* SQUID magnetometer. During the measurements the samples were mounted between two quartz tubes of similar dimensions. This whole configuration was then attached to a sample probe rod. The tubes were cleaned before each measurement to avoid contamination by any magnetic particles. Data were collected, usually in the range of 5 K to 400 K, using a dc magnetic field of 5 Tesla applied along the longest sample dimension. A scan length of 6 cm was used. Sometimes plastic straws were used as sample holder. The background signals were subtracted from the data to obtain the magnetic moment of the sample, from which the molar susceptibilities were obtained. The $\chi(T)$ data for a number of $La_{2-x}Sr_xCu_{1-y}Zn_yO_4$ and Zn-free $YBa_2Cu_3O_{7-\delta}$ compounds are shown in Figs. 8. An important feature of the $\chi(T)$ data for Zn doped La214 is the crossing temperature, $T_{cs}$, which probably occurs due to the dilution effect as Zn substitution removes Cu spins from the $CuO_2$ planes and as well as due to the Zn induced enhancement of $\chi(T)$.

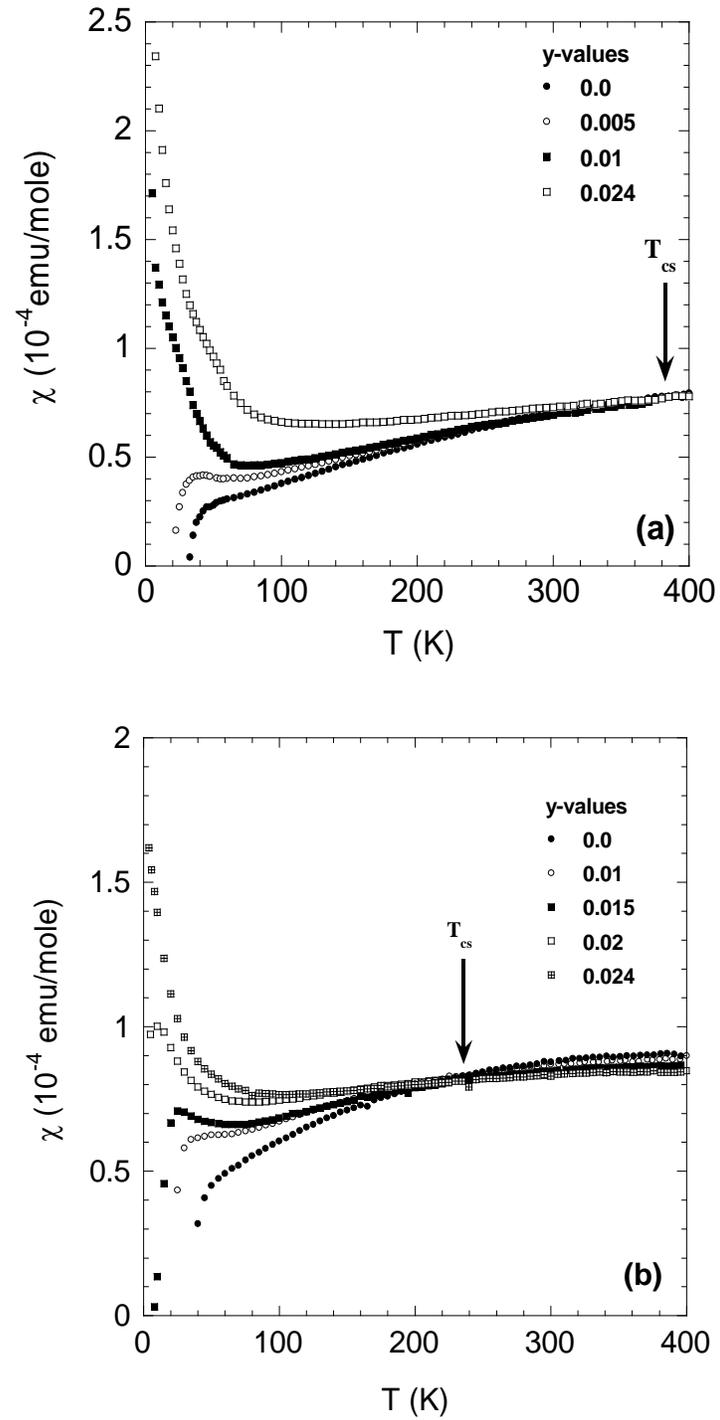

Fig. 8a  $\chi(T)$ for (a) $La_{1.91}Sr_{0.09}Cu_{1-y}Zn_y$ and (b) $La_{1.85}Sr_{0.15}Cu_{1-y}Zn_y$ compounds. $T_{cs}$ marks the temperature at which $\chi(T, y)$ plots cross each other.

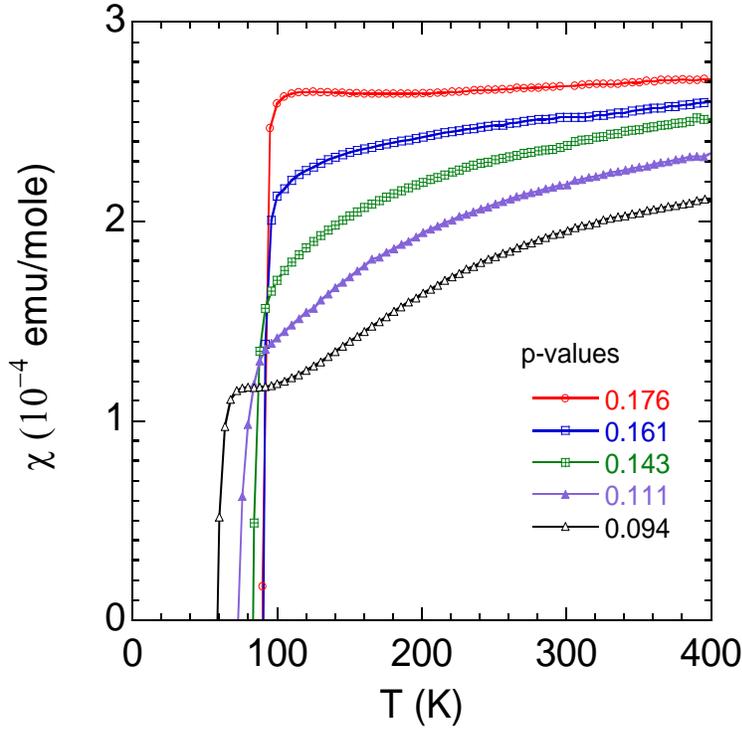

Fig. 8b $\chi(T)$ for $YBa_2Cu_3O_{7-\delta}$ compounds. The hole contents are shown in the plot.

In the subsequent subsections we have analyzed the ACS, $\rho(T)$, and $\chi(T)$ data and investigated the effects of Zn on $T_c$ and the characteristic PG energy scale, respectively, as the in-plane hole content varies in $La_{2-x}Sr_xCu_{1-y}Zn_yO_4$ and $YBa_2(Cu_{1-y}Zn_y)_3O_{7-\delta}$.

**(a) Effect of Zn on $T_c$**

We have plotted the $T_c(p, y = 0)$ and $dT_c(p)/dy$ results for $La_{2-x}Sr_xCu_{1-y}Zn_yO_4$ and $YBa_2(Cu_{1-y}Zn_y)_3O_{7-\delta}$ in Figs. 9a and 9b, respectively. For direct comparison we have shown $dT_c(p)/dy$ for $La_{2-x}Sr_xCu_{1-y}Zn_yO_4$ and $YBa_2(Cu_{1-y}Zn_y)_3O_{7-\delta}$ together in Fig. 10. Figs. 9a and 9b show clearly the strongly $p$-dependent nature of $dT_c(p)/dy$. It is seen that the magnitude of $dT_c(p)/dy$, except near $p \sim 0.125$, decreases systematically with increasing $p$ in the UD to optimally doped region (passing through a minimum in the OD near $p \sim 0.20$ and increases slowly again for further overdoping for $La_{2-x}Sr_xCu_{1-y}Zn_yO_4$). Size of this anomaly is significantly less pronounced for the $YBa_2(Cu_{1-y}Zn_y)_3O_{7-\delta}$ compounds.

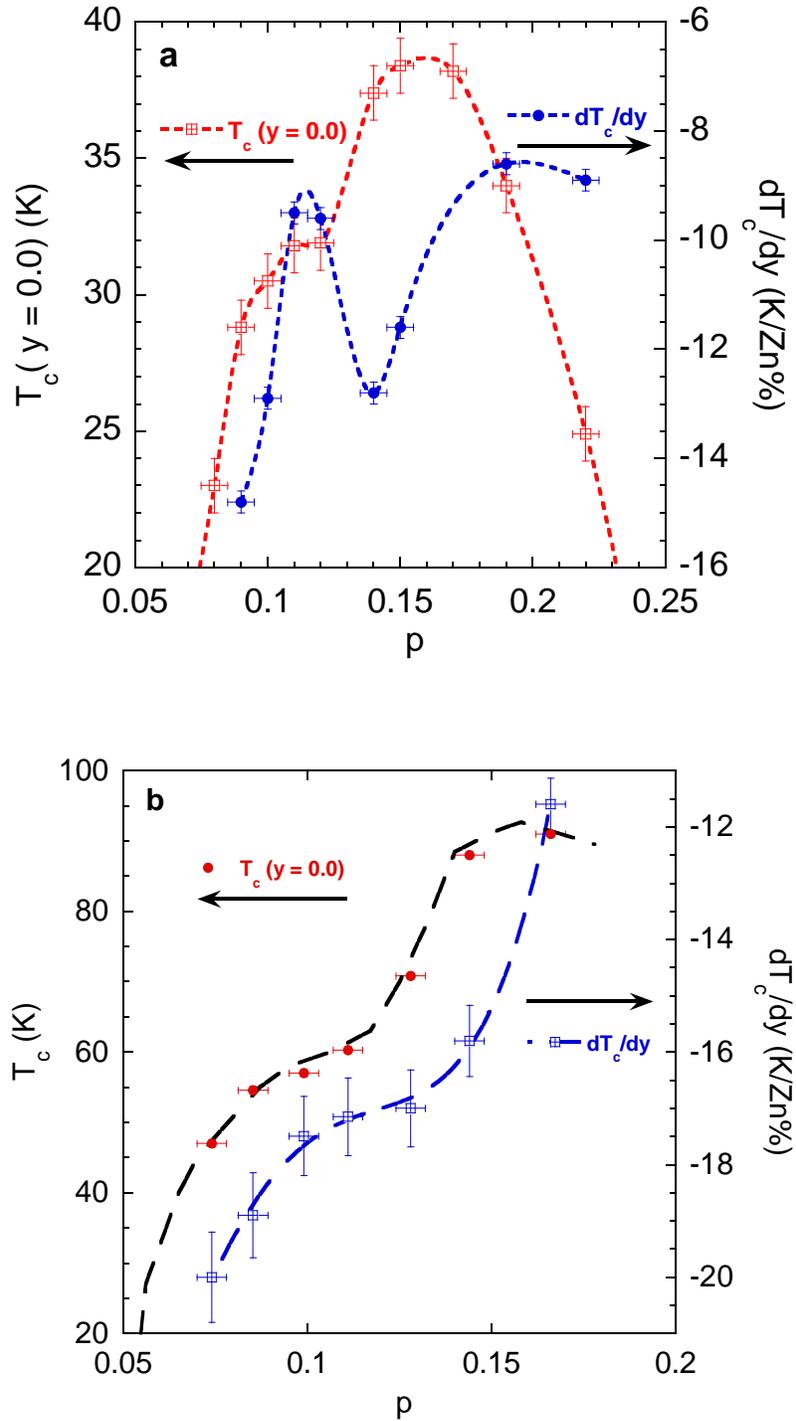

Fig. 9  $T_c(y = 0)$ and $dT_c/dy$ versus $p$ for (a) $La_{2-x}Sr_xCu_{1-y}Zn_yO_4$ and (b) $YBa_2(Cu_{1-y}Zn_y)_3O_{7-\delta}$ compounds. The thick dotted/dashed lines are drawn as guides to the eyes.

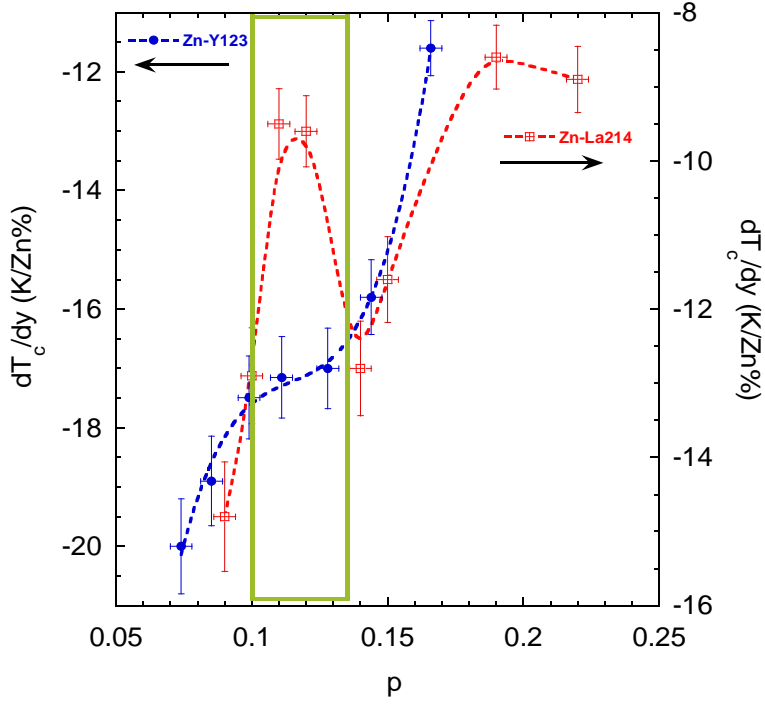

Fig. 10 $dT_c/dy$ versus $p$ for $La_{2-x}Sr_xCu_{1-y}Zn_yO_4$ and $YBa_2(Cu_{1-y}Zn_y)_3O_{7-\delta}$ samples shown together. The boxed part encloses the anomalous region near the $1/8^{th}$ doping.

The anomalous decrease of $dT_c(p)/dy$ in the vicinity of the $1/8^{th}$ doping implies that the effect of Zn on degrading $T_c(p)$ becomes less effective when stripe ordering is at its strongest. In general, the non-magnetic and iso-valent Zn induced rate of suppression of $T_c$ in cuprates can be reasonably well-understood within the framework of strong potential (unitary) scattering of Cooper pairs with $dx^2$-$y^2$ order parameter. Evidence for such pair-breaking scattering by Zn was clearly found from the phase shift analysis of the pioneering STM results obtained by Pan *et al.* [33]. In this unitary scattering formalism, the $p$-dependent variation of $dT_c/dy$ mainly arises from the $p$-dependence of the PG energy scale [34, 35], as the scattering rate is inversely proportional to the thermal average of the electronic density of states at the Fermi-level. Such a picture can address the observed behavior of the $dT_c(p)/dy$ in the UD $La_{2-x}Sr_xCu_{1-y}Zn_yO_4$ and $YBa_2(Cu_{1-y}Zn_y)_3O_{7-\delta}$ and slightly OD compounds except around the $1/8^{th}$ anomaly. On the hand, the slow increment in the magnitude of $dT_c(p)/dy$ in the OD side above $p \sim$ 0.19, is indicative of a fundamental change in the electronic ground state of different origin [36, 37] where the 3D [34] Fermi-liquid description becomes more applicable and a tendency towards metal-superconducting phase separation starts to grow [34] which reduces the SC volume fraction. The size of the anomaly observed in $dT_c(p)/dy$ near $p \sim 0.125$ indicates that a simple unitary scattering scenario with a PG in the EDOS is not adequate and a different additional competing mechanism is at play. Considering that static stripe order always reduces $T_c$ and the superfluid density is also suppressed in the vicinity of $p \sim 0.12$ [38], it is rather surprising to find Zn becoming less effective in reducing $T_c$ in samples where superconductivity is already substantially weakened. This observation becomes even more counterintuitive when one takes into account of the Zn induced pinning of the stripe fluctuations [39 – 41]. The pinning or slowing down of spin/charge fluctuations by Zn can be attributed to the enhancement of the AFM correlations, carrier localization, or to the increase in the stripe inertia. Irrespective of the precise mechanism, as static charge/spin ordering degrades superconductivity, Zn substitution should become more effective in reducing $T_c$ near the $1/8^{th}$ doping. Exactly the opposite effect is found here experimentally. The possible reasons for this might be the followings (i) since spin/charge ordering near $p \sim 0.125$ is already static or quasi-static in the pure compound, Zn substitution plays no significant role in further pinning in this region. In this proposal Zn substitution in cuprates with $p$ close to the $1/8^{th}$ value is "*wasted*" to some extent. This supports the theoretical proposal by Smith *et al*. [42] which describes stripe-pinning as primary mechanism for degradation of $T_c$ due to Zn. (ii) Zn substitution destroys the integrity of the static stripe order. For randomly substituted Zn atoms in the Cu sites, part of the Zn in the hole-rich regions will lead to carrier localization and the other part will replace the antiferromagnetically correlated Cu spins in the hole-poor regions, thereby creating spin vacancies. In such a situation with large number of spin vacancies in the $CuO_2$ planes, a hole from a neighboring domain can hop inside the spin ordered region. This process can destroy the stripe order itself [37, 43]. Within this proposal Zn becomes less effective in reducing $T_c$ because the static stripe order is weakened and consequently superconductivity itself somewhat *strengthened*. The tendency towards stripe formation is found to be stronger in La214 compounds compared to Y123. Incommensurate low-energy spin fluctuations are present in Y123 over a wide range of $p$ [6, 7] but there is no concrete evidence of static charge ordering in this material. This offers an explanation for the reduced size of the anomaly in $dT_c(p)/dy$ around $p$ = 0.125 in YBa$_2$(Cu$_{1-y}$Zn$_y$)$_3$O$_{7-\delta}$ as compared to that found for La$_{2-x}$Sr$_x$Cu$_{1-y}$Zn$_y$O$_4$.

**(b) Effect of hole content and Zn on the pseudogap energy scale: resistive features**

Fig. 7 shows the evolution of the resistivity with hole content. $\rho(T)$ for the optimally and underdoped samples gradually develops an increasingly negative curvature with decreasing $p$. The slope of $\rho(T)$ increases systematically as $p$ is reduced. The residual resistivity also increases monotonically with increasing underdoping. The variation of $\rho(T)$ with $p$ provides us with a way for establishing the $T$-$p$ phase diagram of high-$T_c$ superconductors and can give measures of $T^*(p)$, the characteristic pseudogap temperature. Conventionally, $T^*$ is defined as the temperature at which $\rho(T)$ starts to decrease at a faster rate than from its high-$T$ linear behavior [22, 44]. In this study two somewhat equivalent methods have been employed to locate $T^*$. Fig. 11 illustrates theses procedures. It is seen that plots of $d\rho(T)/dT$ versus $T$ and $[\rho(T) - \rho_{LF}]$ versus $T$ yield very similar $T^*$ values (within ± 5 K). Here, $\rho_{LF}$ is a linear fit of the form $\rho_{LF} = b + cT$, in the high-temperature region of the $\rho(T)$. As a third and more objective procedure, it is possible to scale the $\rho(T)$ data by using $T^*/T$ as the scaling parameter. In a previous article [29] we have shown the results of such scaling for large number samples with $p$ up to 0.20. Such scaling also yields $T^*$ values that are in excellent agreement with those obtained by the other two methods as shown in Fig. 11. Unlike the hole content in the CuO$_2$ plane, in-plane Zn content, at least up to a certain level, does not affect the PG feature in $\rho(T)$ significantly. We have shown a typical example in Fig. 12 (taken from refs. [22, 29]). The fact that $T^*(p)$ is insensitive to Zn whereas $T_c$ itself is highly degraded by it, is indicative that SC and PG correlations might be unrelated. At this point it should be mentioned that resistivity measurement is a bulk probe and for intrinsically disordered cuprate superconductors where the hole content is inhomogeneous over microscopic length scales, $\rho(T)$ only gives an average estimate of the PG energy scale. In fact Zn doping transfers QP spectral weight from high energy to inside the PG on a local scale.

Far from the Zn site the PG sustains its bulk value [45]. Therefore, the onset of the downturn in the resistivity does not shift up to a certain level of Zn substitution, but the downturn becomes weakened and even an additional upturn in $\rho(T)$ can appear at lower temperatures as a precursor to superconductor-insulator transition, depending on hole content. The phonon structures of the various cuprates are fairly insensitive to the hole content, thus it is reasonable to assume that the systematic change in $\rho(T, p)$ is primarily due to electronic correlations. This is also indicative that non-phonon mechanisms dominate in hole doped cuprates. The details of the $\rho(T)$ features are unconventional and a complete understanding is still lacking. An electronic phase diagram dominated by quantum critical point and a PG in the QP spectral density might provide us with a possible framework to address the temperature and hole content dependent resistivity for cuprates [46].

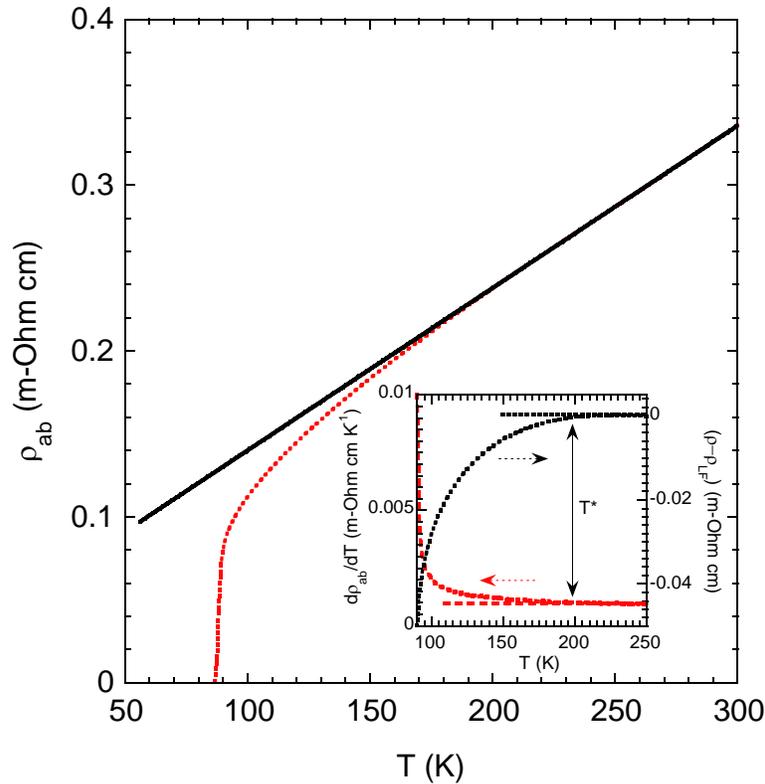

Fig. 11 Extraction of $T^*$ from the $\rho(T)$ data for an underdoped YBa$_2$Cu$_3$O$_{7-\delta}$ thin film with $p$ = 0.140 ± 0.004. Main panel: $\rho_{ab}(T)$ data (red dotted curve) and its high-$T$ linear fit (full black straight line). The inset shows $T^*$ at the high-$T$ onset of the downturn in $\rho_{ab}(T)$.

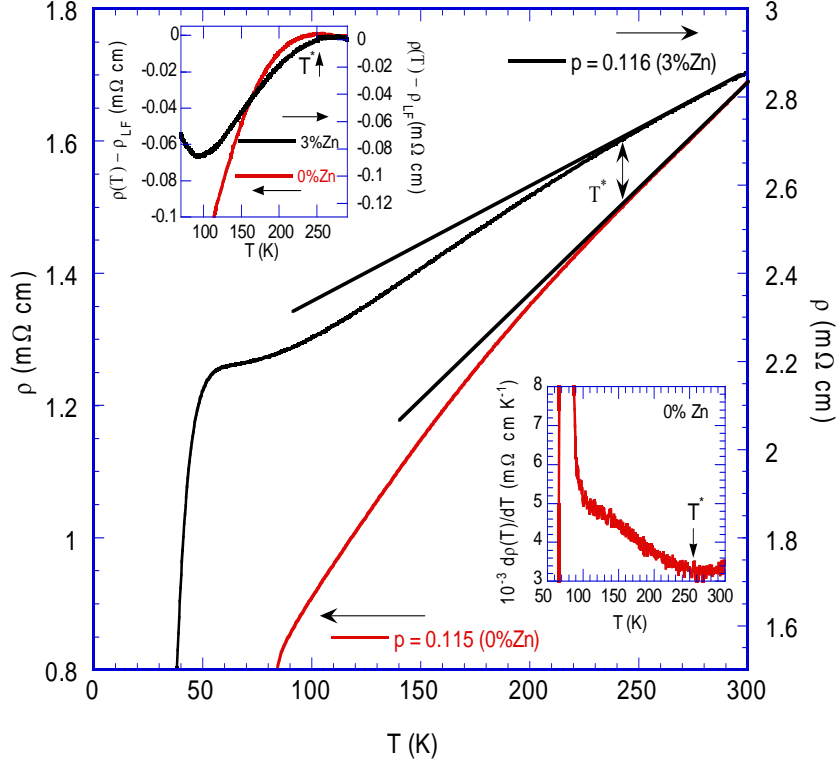

Fig. 12 Zn independence of $T^*$ for underdoped $Y_{0.80}Ca_{0.20}Ba_2(Cu_{1-y}Zn_y)_3O_{7-\delta}$ compounds. $p$-values are shown and are accurate within $\pm 0.004$ (taken from refs. [22, 29]). Notice the low-$T$ upturn in $\rho(T)$ for the Zn substituted compound (see text for details).

**(c) Extraction of the PG energy scale from the bulk magnetic susceptibility**

A systematic study of the temperature dependent uniform magnetic susceptibility, $\chi(T)$, over a wide range of hole concentration can yield valuable information regarding the origin of PG [47, 48]. The most important information that can be extracted from the analysis of $\chi(T, p)$ is about the temperature and $p$ dependencies of EDOS. As EDOS as a function of energy lies at the heart of any problem associated with PG, $\chi(T, p)$ is indeed a powerful tool to study various $p$-dependent features of this phenomenon. It is in fact the intrinsic spin part (the Pauli spin susceptibility, $\chi_{spin}$) of $\chi(T, p)$ that represents the QP spectral weight near the Fermi energy. Pauli spin susceptibility arises from the coupling of intrinsic spins of the mobile charge carriers with the applied magnetic field and, for ordinary Fermi-liquids with $\varepsilon_F \gg k_B T$, can be approximated as

$$\chi_{spin} = \mu_B^2 N(\varepsilon_F) \qquad (1)$$

where $\mu_B$ is Bohr magneton and $N(\varepsilon_F)$ is the EDOS at the Fermi level. Therefore, $\chi_{spin}$ at a particular temperature $T$ represents the thermal average value $\langle N(\varepsilon)\rangle$ over an energy window of width $\sim \varepsilon_F \pm 2k_BT$. The above expression, although derived for Fermi-liquids, holds for underdoped cuprates where strong electronic correlations are present and integrity of QP description has been questioned. We address this issue in the subsequent sections. Perhaps the most direct bulk probe of the EDOS is the electronic heat capacity measurements. The electronic specific heat coefficient, $\gamma_{el}(T)$, for Fermi-liquids gives a direct measure of $N(\varepsilon)$ around the chemical potential. The zero magnetic field electronic entropy, $S_{el}(T)$, obtained from $\int \gamma_{el}(T)dT$ shows a striking qualitative and quantitative resemblance with the $\chi(T)T$ data for various cuprate superconductors including La214 and Y123 [48]. This shows that $\chi(T)$ probes the same underlying EDOS as probed by the heat capacity measurements. The electronic entropy measures the number of thermal excitations over the entire electronic spectrum (spin and charge over all wave vectors) whilst $\chi(T)T$ provides complementary information for the $q = 0$ spin spectrum. In fact $\chi_{spin}(T)k_BT/\mu_B^2$ gives a measure of the number of thermally excited spin excitations at a particular temperature $T$ [48]. Eq. (1) relates electronic QP EDOS to the Pauli spin susceptibility. These QPs carry both charge and spin. These also assure that the PG observed in the spin sector via $\chi_{spin}(T)$ is the same as that revealed when thermally excited QPs are probed through electronic heat capacity measurements.

From the detailed analysis of the heat capacity data it was found [48, 49] that a *states non-conserving* triangular gap pinned at the chemical potential describes the experimental features of $S_{el}(T)$ and $\gamma_{el}(T)$ quite well. Following this lead we have modeled the low-energy EDOS as follows:

$$
\begin{aligned}
N(\varepsilon) &= N_0|\varepsilon/\varepsilon_g| & \text{for} \quad &\varepsilon \leq \varepsilon_g \\
&= N_0 & \text{for} \quad &\varepsilon > \varepsilon_g
\end{aligned}
$$

$$(2)$$

where $N_0$ is the flat EDOS outside the PG region. Using Eq. (2) in the usual expression for the Pauli spin susceptibility given by-

$$\chi(T) = \mu_B^2 \int (-\partial f(\varepsilon, T)/\partial \varepsilon) N(\varepsilon) d\varepsilon \qquad (3)$$

(where $f(\varepsilon, T)$ is the Fermi function) yields,

$$\chi(T) = \mu_B^2 N_0 [1 - (2k_B T/\varepsilon_g) \ln\{\cosh(\varepsilon/2k_B T)\}] \qquad (4)$$

Here, the constant $\mu_B^2 N_0$ largely takes into account of all the $T$-independent contributions to the bulk magnetic susceptibility. We have used such a model PG to analyze the $\chi(T)$ data previously [50, 51]. A similar triangular PG model was also employed to study the STM results for Zn substituted Bi2212 [52]. Experimental indications favoring *states non-conserving* nature of the PG were also found from tunneling [53] and NMR studies [54]. It is worth remembering that the experimentally measured $\chi(T)$ consists of several contributions of different physical origins and not all of them are directly linked to the EDOS. For example, the contributions to $\chi(T)$ due to *Larmor* or the core susceptibility and the *Van Vleck* susceptibility are expected to be small and $p$- and $T$-independent over the experimental temperature range [55, 56]. Whereas the *Landau* diamagnetic susceptibility has the same $T$-dependence as $\chi_{spin}(T)$ (in ordinary Fermi-liquids with an almost constant EDOS near $\varepsilon_F$, both are approximately $T$-independent). For cuprates *Landau* diamagnetic contribution is only from 2% to 5% of the $\chi_{spin}(T)$ in magnitude [55]. Therefore, a fit of the experimental $\chi(T)$ using Eq. (4) would overestimate $N_0$ to some extent, keeping the value of $\varepsilon_g$ unaffected at a particular hole content. A significant but extrinsic source of error in determining $\varepsilon_g$ from such fit is the possible presence of magnetic impurity phases. For example, small amount of paramagnetic impurity, mimicking a *gap filling* feature, can largely reduce the extracted value of the $\varepsilon_g$. In this study all the samples were found to be phase-pure within the resolution of the XRD machine. Results of fits to $\chi(T)$ down to $\sim T_c + 30$ K are shown in Figs. 13. The lower temperature limit has been set to avoid regions with strong diamagnetic superconducting fluctuations. The quality of fits for Y123 is excellent. For La214 with $x < 0.18$ it is quite good but perhaps a little inferior to the ones for Y123. The possible reasons might be that intrinsic disorders in La214 due to Sr substitution and the enhanced tendency towards static/dynamic (depending on the Sr content) stripe correlations present in this compound. The simple PG model used here would not be able to account for such complexities which should affect the EDOS in non-trivial fashion. For the $x = 0.18$ compound the fit is not as good as the two other UD samples but the extracted value of the $\varepsilon_g/k_B$ agrees very well with those obtained from the analysis of the zero (magnetic) field heat capacity data by Loram *et al* [48, 49]. All the fits yield almost identical values for $N_0$ irrespective of $p$ (for Y123, $\mu_B^2 N_0 = 2.75 \pm 0.10$ and for La214 it is $1.05 \pm 0.08$). In Fig. 13b we have also shown fits for the 19% and 22% Sr compounds. The qualities of these fits are quite poor. This is probably due to the absence of PG in these two compounds. In such a case Eq. (4) is not valid any longer. For example, the $\varepsilon_g/k_B$-value obtained for the $x = 0.22$ compound is ~ -300 K. The *unphysical* negative value is indicative of a triangular-like enhancement of the EDOS near the chemical potential rather than a triangular depletion due to the PG. Such gradually enhanced EDOS in the OD La214 have been reported by other studies [57] and might well be related to the van-Hove singularity (vHS) scenario. In the vHS scenario, the EDOS at the Fermi level exhibits a logarithmic singularity and affects a number of electronic properties, including $T_c$, in non-trivial manners.

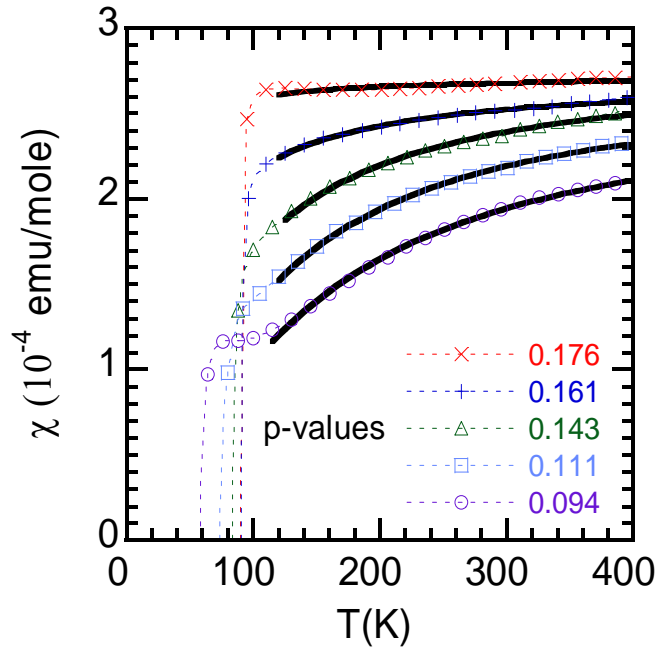

Fig. 13a $\chi(T, p)$ of pure Y123. Thick full lines show the fits to Eq. (4).

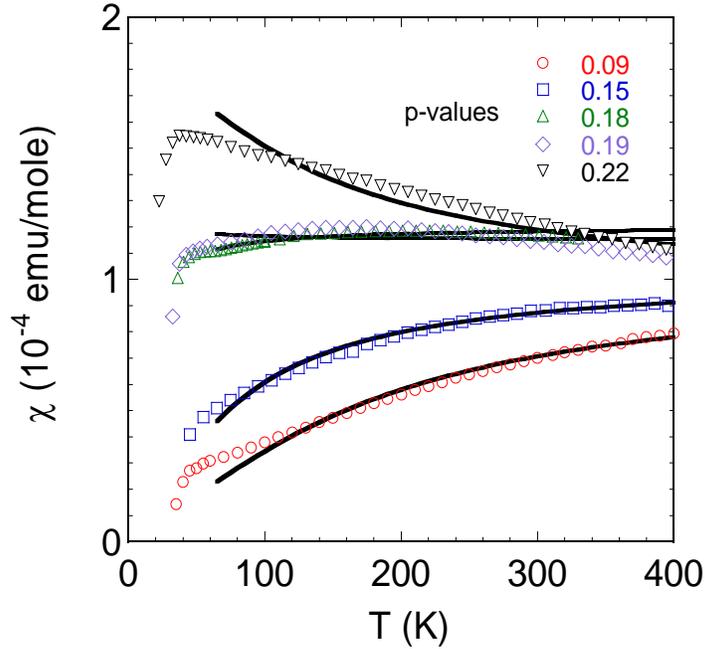

Fig. 13b $\chi(T, p)$ of pure La214. Thick full lines show the fits to Eq. (4).

We have shown the $\varepsilon_g(p)/k_B$-values extracted from the fits to $\chi(T, p)$ and experimental $T_c(p)$ values of Y123 and La214 in Fig. 14. Despite the widely different $T_c(p)$, $\varepsilon_g(p)/k_B$-values are nearly identical for Y123 and La214. Also for both the families $\varepsilon_g(p)/k_B$ vanishes at the same hole content, $p \sim 0.19$, as found in some previous studies using different experimental probes [1, 44, 48, 49, 58].

An interesting result of non-magnetic Zn substitution in hole doped cuprates is the impurity-induced magnetism. As mentioned earlier, various experiments [17 – 20] gave indications that Zn gives rise to a local magnetic moment-like feature on the four nearest neighbor Cu sites in the $CuO_2$ planes. This apparent magnetic behavior is evident from the appearance of the Curie-like term in the bulk magnetic susceptibility, $\chi(T)$, in the normal state [17, 18]. In this study we have modeled the bulk magnetic susceptibility for the Zn doped La214 and Y123 mainly in terms of a depleted EDOS at low-energy due to the PG and a Zn induced enhancement resembling a Curie contribution. Therefore, the modified form of Eq. (4) becomes-

$$\chi(T) = \mu_B^2 N_0[1 - (2k_BT/\varepsilon_g)ln\{cosh(\varepsilon/2k_BT)\}] + C(y)/T \qquad (5)$$

$C(y)$ is a Zn dependent term resembling to a Curie constant as in ordinary paramagnets.

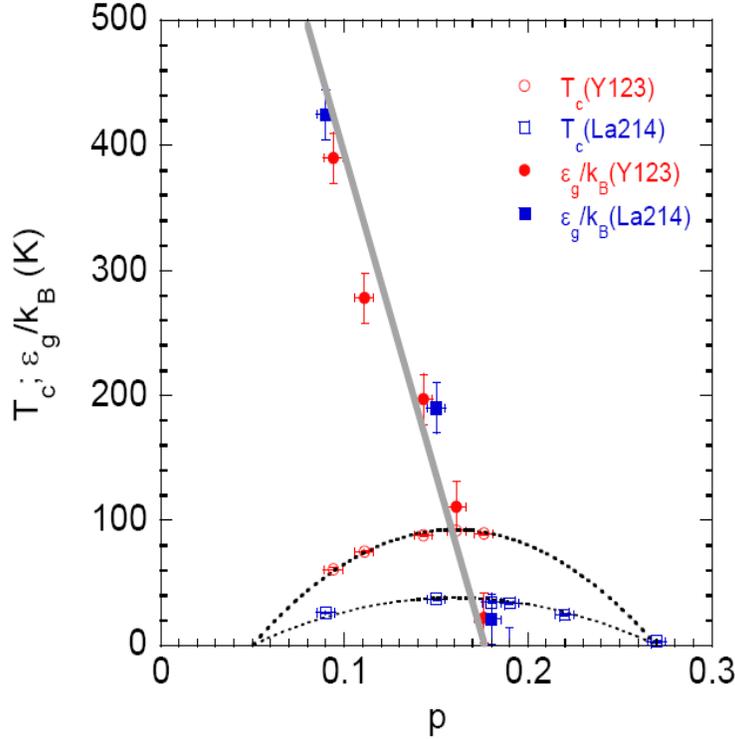

Fig. 14  $T_c(p)$ and $\varepsilon_g(p)/k_B$ for pure Y123 and La214. The thick full line is drawn as a guide to the eyes.

Results of fitting of $\chi(T)$ data to Eq. (5) for Zn substituted La214 are shown in Figs. 15. $C(y)$ values extracted from these fits to the above mentioned equation show a linear trend and from the slopes of these fitted straight lines (see Fig. 16), the effective number of Bohr magnetons per Zn, $p_{eff}$/Zn, were calculated using the relation $C/y = (N_A\mu_B^2 p_{eff}^2)/3k_B$, here $N_A$ is Avogadro's number. This analysis yields $p_{eff}$/Zn = 0.91 ± 0.08 and 0.78 ± 0.07 (in the units of $\mu_B$) for the 9% and 15% Sr substituted compounds, respectively. These values are in good agreement with those obtained by earlier studies (for different families of cuprates) at similar values of hole contents [59 – 61], using different methods of analysis. PG values were once again found to be fairly insensitive to the level of Zn substitution at a given $p$. Values of $\varepsilon_g/k_B$, extracted from the analysis of $\chi(T)$ data with different Zn contents are shown in Fig. 17.

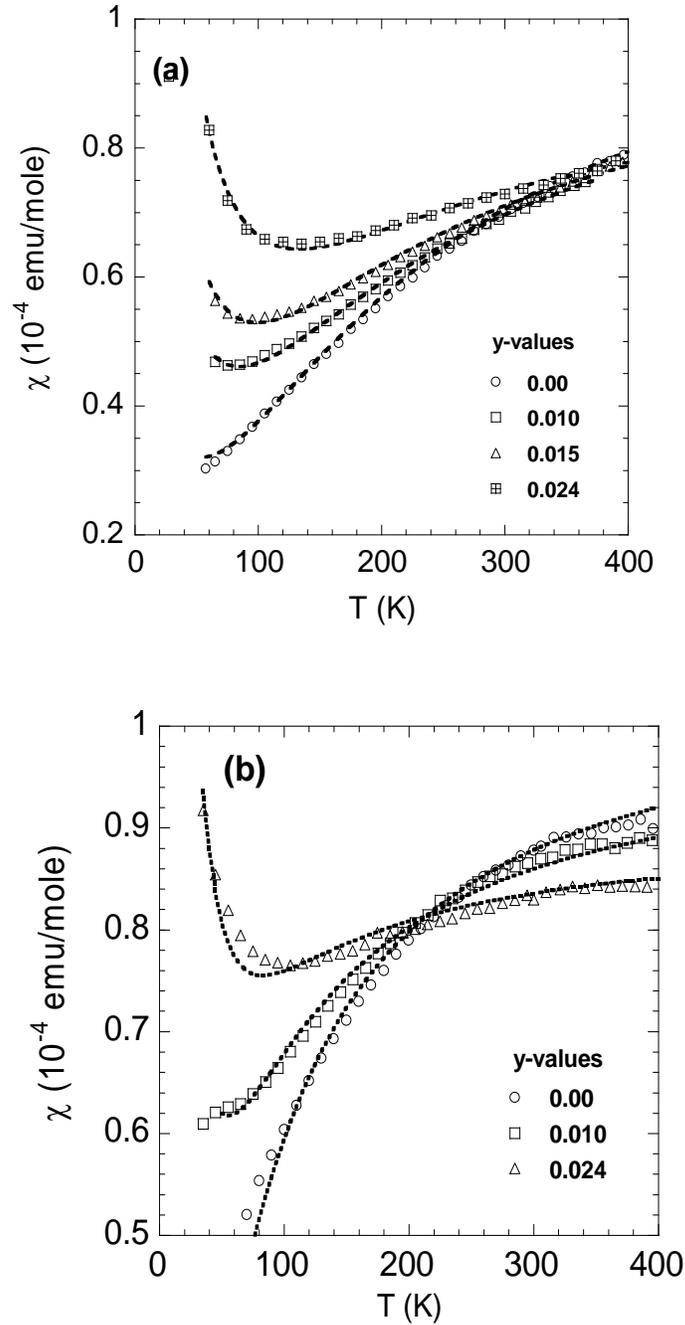

Fig. 15 Representative $\chi(T)$ fits (dashed lines) for (a) $La_{1.91}Sr_{0.09}Cu_{1-y}Zn_y$ and (b) $La_{1.85}Sr_{0.15}Cu_{1-y}Zn_y$ compounds. Symbols show experimental data points.

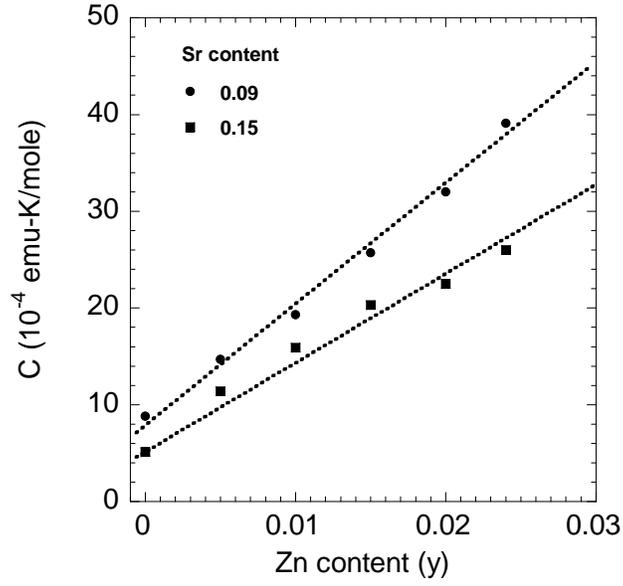

Fig. 16  $C$ versus Zn content ($y$) for $La_{1.91}Sr_{0.09}Cu_{1-y}Zn_y$ and $La_{1.85}Sr_{0.15}Cu_{1-y}Zn_y$ samples. The dashed straight lines are drawn as guides to the eyes.

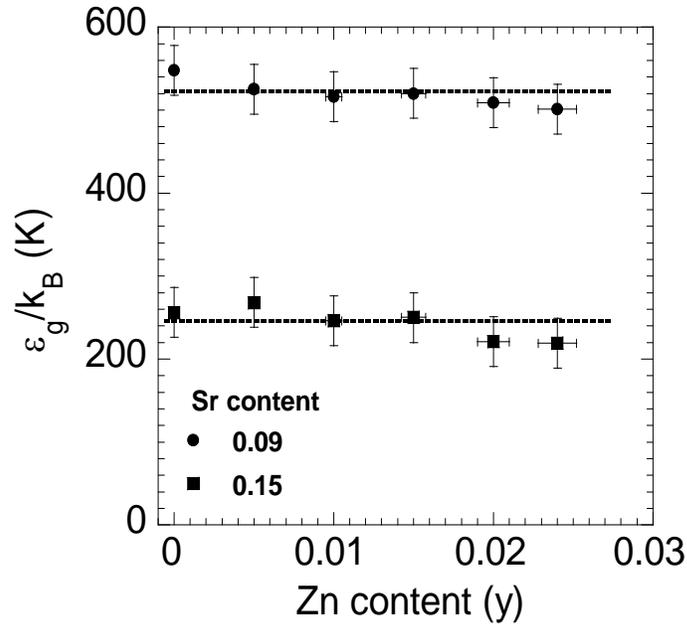

Fig. 17  The characteristic pseudogap temperature ($\equiv \varepsilon_g/k_B$) versus Zn content for $La_{1.91}Sr_{0.09}Cu_{1-y}Zn_y$ and $La_{1.85}Sr_{0.15}Cu_{1-y}Zn_y$. The dashed straight lines are drawn as guides to the eyes.

It is of some interest that fits to Eq. (5) yield small but finite values of $C(y)$ even for Zn free La214. This apparent Curie constant in the Zn-free samples might be related to an intrinsic disordering effect present in $La_{2-x}Sr_xCuO_4$ compounds. As can be seen from Fig. 16, the value of $C(y = 0)$ lies on the same line drawn for $C(y)$ at a particular level of Sr content. This suggests that an intrinsic amount of disorder, capable of giving rise to an enhanced magnetic response at low-$T$, is present in all the $La_{2-x}Sr_xCuO_4$ compounds used in this study. Sr substitution introduces unavoidable disordering effect in the neighboring $CuO_2$ plane. The fact that it has an effect similar to that due to Zn doping is not surprising. For example, Albenque *et al.* [62] demonstrated that (non-magnetic) in-plane defects induced by electron irradiation can give rise to magnetism and pair-breaking in high-$T_c$ cuprates, just as Zn does. The effect of Zn on $\chi(T)$ seems to follow closely the magnitude of the PG energy scale. Higher the value of $\varepsilon_g$, higher is the value of $p_{eff}$/Zn.

## 4. Discussion and Conclusions

The effect of Zn on the superconducting transition temperature in cuprates has been investigated by a large number of previous studies, both experimentally and theoretically [15, 16, 34 – 37, 63 – 65]. Most of these previous studies did not address the anomalous feature in $dT_c(p)/dy$ in the vicinity of the 1/8[th] doping. In Section 3(a) we have described the possible scenarios that can lead to an anomalous decrease in the magnitude of $dT_c(p)/dy$ when $p \sim 0.125$. To develop a coherent theoretical scheme, one needs to know the microscopic details regarding the interplay between spin/charge stripe and SC correlations, which is still unclear at this stage. At this point, we would like to mention that some of the previous work on Zn doped cuprates reported an enhanced rate of suppression of $T_c$ at around the 1/8[th] doping [63, 64], in complete contrast with the findings reported here. We believe this contrasting behavior resulted from sample related issues and also from the level of Zn substitution. For example, a careful inspection of the susceptibility and resistivity data presented in ref. [65] indicates that the samples might contain unintentional additional disorders at various hole contents. It is important to use a relatively high level of Zn substitution to calculate $dT_c(p)/dy$, since a low level of Zn doping invariably amplifies the magnitude of $dT_c(p)/dy$. This is due to the fact that to disrupt stripe order effectively, at least a moderate amount of Zn is required inside the $CuO_2$ planes. Previously, it was also suggested that the 60 K plateau in the $T_c(\delta)$ of Y123 arises from the ordering of the doped oxygen atoms in the $CuO_{1-\delta}$ chain sites [66]. But careful studies on Ca doped Y123 [29, 67] and pure Y123 [68] compounds provided with strong evidences that the 60 K plateau actually originates from the hole content in the $CuO_2$ planes irrespective of the oxygen arrangement in the $CuO_{1-\delta}$ chains. These studies [67 – 69] also linked the presence of the 60 K plateau intimately to the $1/8^{th}$ hole content. It is also important to mention that the iso-valent Zn mainly replaced the in-plane Cu atoms in all our samples. Zn does not show a tendency to reside in the chain Cu sites [70], at least up to $y = 0.07$ [23].

As far as the resistivity is concerned, the characteristic PG temperatures were found to be insensitive either to the crystalline states of the experimental sample or to the levels of Zn substitution. The former indicates that in polycrystalline compounds resistivity is primarily dominated by the charge conduction within the $ab$ plane, while the later indicates that Zn affects the pairing gap below $T_c$ and the PG at high-$T$ in fundamentally different fashion. The onset of the PG feature in $\rho(T)$ is gradual and rather weak (it is even weaker in the Zn doped compounds). Such a crossover like behavior does not support the various theoretical proposals where the development of the PG has been discussed in terms of novel electronic phases and ordered states [71]. A possible explanation might be due to the presence of intrinsic imperfections in all cuprate superconductors, which could wash away the expected sharp features associated with phase transitions, otherwise revealed via various transport and magnetic measurements. Zn doping induces an additional upturn in $\rho(T)$ of UD cuprates. It could be related to the in-(pseudo)gap states created by Zn or to the spin degrees of freedom added by the Zn atoms. It has been proposed that these spin degrees of freedom play a role in the charge scattering, and induce, for instance, a (Kondo-like) spin flip contribution to the resistivity which changes the nature of the temperature dependence of $\rho(T)$ at low temperatures.

$La_{2-x}Sr_xCu_{1-y}Zn_yO_4$ and $YBa_2(Cu_{1-y}Zn_y)_3O_{7-\delta}$ have significantly different anisotropy factors and superconducting transition temperatures. Despite these differences, analysis of $\chi(T)$ data over a wide range of hole contents and in-plane disorder yielded almost identical values for the characteristic PG energy scale. Within the precursor pairing scenario of the PG, strong SC fluctuations due to two-dimensional nature and low superfluid density inhibits the formation of phase coherent cooper pairs at higher temperatures and fluctuating cooper pairs present at these temperatures give rise to the PG in the strong coupling regime. In this picture PG increases with underdoping because of the stronger coupling whereas decreasing superfluid density and increasing anisotropy (both enhancing superconducting fluctuations) decreases $T_c$ (at which phase coherent cooper pairs are formed). As hole content increases, superfluid density increases, anisotropy decreases, and PG temperature comes closer to the phase coherence temperature, *i.e.*, $T_c$. Theoretical calculations have also shown that anisotropy, and therefore, low-dimensionality plays a crucial role in formation of the PG, in the precursor pairing scenario [72]. Our findings here do not support these ideas regarding the origin of the PG. Also, the very facts that $\varepsilon_g(p)/k_B$ goes below $T_c(p)$ in the slightly OD side (*i.e.*, persists inside the SC dome) and vanishes at $p \sim 0.19$ cannot be accommodated in theories involving fluctuating cooper pairs where the PG temperature merges with $T_c$ in the OD side as pair formation and phase coherence temperatures approach each other. The extrapolated $\varepsilon_g(p)/k_B \sim 1000$ K at $p = 0$ for Y123 and is about 20% higher for La214. These extrapolated values are quite close to the values of the antiferromagnetic (AFM) exchange energies, $J_{ex}$, for the parent compounds (YBa$_2$Cu$_3$O$_6$ and La$_2$CuO$_4$). It is worth mentioning that the magnitude of $T^*(p)$ found from the resistivity data is always somewhat lower than $\varepsilon_g(p)/k_B$ (extracted from $\chi(T)$) at the same hole content. This is not surprising because different experiments probe the energy window centered at the chemical potential in different fashions. It is the nature of the evolution of the characteristic PG energy/temperature scale with hole content that carries the information regarding its interplay with superconductivity, rather than its exact magnitude.

The fact that a states non-conserving PG fits the experimental data satisfactorily, has important possible theoretical consequences. For example, this may imply that the formation of PG transfers QP spectral weight to very high energies (where the lost states in the PG are recovered) or PG arises from some exotic order unlike spin or charge density waves where QP states are conserved. There are several closely related scenarios for the pseudogap invoking some unconventional type of order, *e.g.*, *d*-density wave [73] and orbital current phases [74]. In certain theoretical versions of the orbital current phase [74], it was proposed that this breaks the time-reversal symmetry through spontaneously ordered current loops in the O-Cu-O plaquettes without ever breaking the translational symmetry. Such ordering can lead to a PG without any singularity in the heat capacity (which occurs in general when a second-order thermodynamic phase transition takes place). Polarized neutron diffraction experiments [75] showed indications of a novel magnetic order in UD Y123, consistent with the one predicted by the orbital current phase in the PG region [75].

The extracted Zn induced magnetic moments showed a close correlation with the magnitude of the PG energy scale. Irrespective of the origin of this impurity induced magnetism, it appears that a short-range AFM correlation is essential. This links the origin of the pseudogap to the short-range AFM correlations as suggested in a number of previous publications [1, 25, 48, 58, 76 – 78].

To summarize, we have investigated the effects of Zn substitution on the superconducting transition temperature, in-plane dc resistive features, and bulk magnetic susceptibilities in a series of $La_{2-x}Sr_xCu_{1-y}Zn_yO_4$ and $YBa_2(Cu_{1-y}Zn_y)_3O_{7-\delta}$ compounds over a wide range of hole contents. Both $T^*(p)$ and $\varepsilon_g(p)/k_B$ decrease almost linearly with increasing hole content without any noticeable feature around $p \sim 0.125$. On the other hand both $T_c(p)$ and $dT_c(p)/dy$ show strongly non-monotonic features in the vicinity of the $1/8^{th}$ doping. All these facts, together with the observation that Zn degrades $T_c$ most effectively but has almost no effect on the PG energy scale, are indicative that spin/charge ordering, PG correlations, and superconductivity have different electronic origins and probably competes with one another over a significant region in the $T$-$p$ phase diagram.

**Acknowledgements**

The authors would like to thank Dr. J. R. Cooper (University of Cambridge, UK), Dr. J. W. Loram (University of Cambridge, UK), and Prof. J. L. Tallon (Victoria University, Wellington, New Zealand) for their help, thoughtful comments, and suggestions over a long period of time. We would also like to thank Commonwealth Scholarship Commission (UK), Trinity College, Darwin College, the Cambridge Philosophical Society, the Lundgren Fund, the Department of Physics and the IRC in Superconductivity, Cambridge, for funding and providing with the experimental facilities for this work at various stages. SHN thanks the Quantum Matter group, University of Cambridge, for their hospitality. We also thank the AS-ICTP, Trieste, Italy, for the hospitality.

*Corresponding author. E-mail: salehnaqib@yahoo.com